\documentclass[conference]{IEEEtran}
\IEEEoverridecommandlockouts
\usepackage{cite}
\usepackage{amsmath,amssymb,amsfonts}
\usepackage{algorithmic}
\usepackage{graphicx}
\usepackage{textcomp}
\usepackage{xcolor}
\usepackage[hyphens]{url}
\usepackage{hyperref}
\usepackage{multirow}
\usepackage{gensymb}
\usepackage{threeparttable}
\usepackage[font=small,labelfont=bf]{caption}
\usepackage[a4paper, total={184mm,239mm}]{geometry}
\def\BibTeX{{\rm B\kern-.05em{\sc i\kern-.025em b}\kern-.08em
    T\kern-.1667em\lower.7ex\hbox{E}\kern-.125emX}}

\hypersetup{
    colorlinks=true,
    linkcolor=blue,
    filecolor=magenta,      
    urlcolor=cyan,
    citecolor = magenta,
    pdfpagemode=FullScreen,
}

\PassOptionsToPackage{bookmarks={false}}{hyperref}

\begin{document}


\title{
\fontsize{22.1}{30.0}\selectfont{\textsc{H3DFact}: Heterogeneous 3D Integrated CIM for Factorization with Holographic Perceptual Representations}
}
\author{
Zishen Wan$^*$, Che-Kai Liu$^*$\thanks{* These authors contributed equally to this work.}, Mohamed Ibrahim, Hanchen Yang, Samuel Spetalnick, \\Tushar Krishna, Arijit Raychowdhury\\
\textit{School of Electrical and Computer Engineering, Georgia Institute of Technology, GA, USA}\\
\normalsize \{zishenwan, che-kai, mibrahim81, hanchen, sspetalnick3\}@gatech.edu, \{tushar, arijit.raychowdhury\}@ece.gatech.edu
}
\maketitle

\begin{abstract}
Disentangling attributes of various sensory signals is central to human-like perception and reasoning and a critical task for higher-order cognitive and neuro-symbolic AI systems. An elegant approach to represent this intricate factorization is via high-dimensional holographic vectors drawing on brain-inspired vector symbolic architectures. However, holographic factorization involves iterative computation with high-dimensional matrix-vector multiplications and suffers from non-convergence problems.

In this paper, we present \textsc{H3DFact}, a heterogeneous 3D integrated in-memory compute engine capable of efficiently factorizing high-dimensional holographic representations. \textsc{H3DFact} exploits the computation-in-superposition capability of holographic vectors and the intrinsic stochasticity associated with memristive-based 3D compute-in-memory. Evaluated on large-scale factorization and perceptual problems, \textsc{H3DFact} demonstrates superior capability in factorization accuracy and operational capacity by up to five orders of magnitude,
with 5.5$\times$ compute density, 1.2$\times$ energy efficiency improvements, and 5.9$\times$ less silicon footprint compared to iso-capacity 2D designs.


\end{abstract}


\section{Introduction}
\label{sec:intro}

The brain's remarkable ability to reason and comprehend the world relies heavily on its capacity to disentangle sensory attributes. This intricate process involves the factorization of various sensory inputs (e.g., vision, hearing, touch) into distinct perceptual features. This factorization not only aids in perception but also serves as the foundation for higher-order cognition like problem-solving and abstract thinking, thus serving as a crucial component for neuro-symbolic AI~\cite{burak2010bayesian,wan2024towardscog,hersche2023neuro,yang2023neuro,wan2024towards}.

An elegant approach to represent this intricate factorization is via high-dimensional holographic vectors in the context of brain-inspired vector-symbolic architecture~\cite{kleyko2023survey,hersche2024probabilistic}. Each sensory attribute is encoded and processed using a unique holographic vector, thereby creating distinct and separable representations. These representations can be manipulated using a set of algebraic operations. For instance, an object with multiple attributes can be described by the element-wise multiplication of all vectors representing these attributes. The factorization problem in turn is concerned with decomposing a product vector into its constituent attribute vectors. This is a hard combinatorial search problem when dealing with complex attribute structures~\cite{kleyko2022vector}.  


The compositional nature of holographic vector representations gave rise to an efficient factorization algorithm, \emph{resonator network}, that equips with superior ability to bridge cognitive gaps in neuro-symbolic AI, by accepting perceptual representations from a neural network and factorizing them for symbolic reasoning~\cite{frady2020resonator,renner2022neuromorphic}. 
Resonator network is able to perform search in superposition, which allows for simultaneous exploration of a product's constituent elements. This factorization procedure exhibits characteristics akin to dynamic systems, engaging in an iterative computation flow with high-dimensional matrix-vector multiplications (MVMs).
It also relies on stochastic exploration strategies to circumvent the potential limit optimization cycle pitfalls. These features make factorization amenable to computing platforms that enable compute-in-memory (CIM) and are inherently stochastic, such as memristive devices~\cite{yu2021compute,crafton2022improving,liu2022cosime,chang202373}. 


Recently, an in-memory factorizer using the resonator network was proposed~\cite{langenegger2023memory}, where each individual die contains a 2D CIM array to accelerate a specific MVM operation. This approach, however, does not exploit the full potential of CIM; it incurs considerable cost due to the increased silicon area and data communication between different dies in each iteration. Our goal is to achieve highly efficient holographic factorization by capitalizing on the capabilities offered by emerging memory technologies with the integration of multiple heterogeneous arrays in a 3D-stacked configuration~\cite{li2023h3datten,li2023emerging,luo2024h3d}. 


In this paper, we propose \textsc{H3DFact}, the first heterogeneous 3D (H3D) integrated CIM factorizer for high-dimensional holographic vector representations. \textsc{H3DFact} features a hybrid memory design, which integrates analog RRAM computation with digital SRAM components. The RRAM tier is used to efficiently process MVM operations and is designed using a legacy technology node to support relatively high programming voltages. The RRAM's peripheral circuitry, on the other hand, is placed on a separate tier and is integrated with SRAM units using a more advanced node. The integration of these tiers via an H3D configuration leads to improvements in silicon area and energy efficiency. Furthermore, the non-deterministic nature of the RRAM memory elements enhances factorization convergence and operational capacity. Compared to iso-capacity 2D designs, \textsc{H3DFact} demonstrates superior efficiency in terms of performance, power, and area.   



This paper, therefore, makes the following contributions:
\begin{itemize}
    \item We propose the first H3D integrated CIM accelerator, \textsc{H3DFact}, for efficient and scalable factorization of high-dimensional holographic representations.
    \item We present a hybrid-memory design that combines the merits of RRAM computation in legacy nodes (40~nm) and digital-SRAM components in advanced nodes (16~nm).
    \item We demonstrate that \textsc{H3DFact} improves factorization accuracy and operational capacity by up to five orders of magnitude by virtue of inherent stochasticity, with 5.5$\times$ compute density, 1.2$\times$ energy efficiency, and 5.9$\times$ less silicon footprint compared to iso-capacity 2D designs.
\end{itemize}



\section{Background and Motivation}
\label{sec:background}

This section presents high-dimensional vector operations (Sec.~\ref{subsec:HD}) for perceptual encoding and factorization (Sec.~\ref{subsec:resonator}), and motivates the proposed 3D integrated CIM solution designed for factorization (Sec.~\ref{subsec:h3d_motivation}).

\subsection{High-Dimensional Holographic Vector Operations}
\label{subsec:HD}
In high-dimensional holographic vector operations, atomic features and patterns can be encoded using randomly generated vectors (\emph{item vectors}) $x_i\in \{-1,+1\}^D$, where $D$ can be in the range of thousands. Due to the randomness and holographic nature of high-dimensional vectors, item vectors are therefore quasi-orthogonal, i.e., dissimilar, allowing for the disambiguation of the different represented features. These vectors can be manipulated using the following operations~\cite{kanerva2009hyperdimensional}: (1) element-wise multiplication ($\odot$), which can be used for ``binding'' item vectors to create a product and also for ``unbinding'' a product to retrieve item vectors; (2) element-wise addition ($[+]$), which computes the superposition of multiple products; (3) permutation ($\rho$), which changes the ordering of vector elements to capture the sequence of the feature. 

\subsection{Factorization \& Resonator Network}
\label{subsec:resonator}
\begin{figure}
    \centering
    \includegraphics[width=\linewidth]{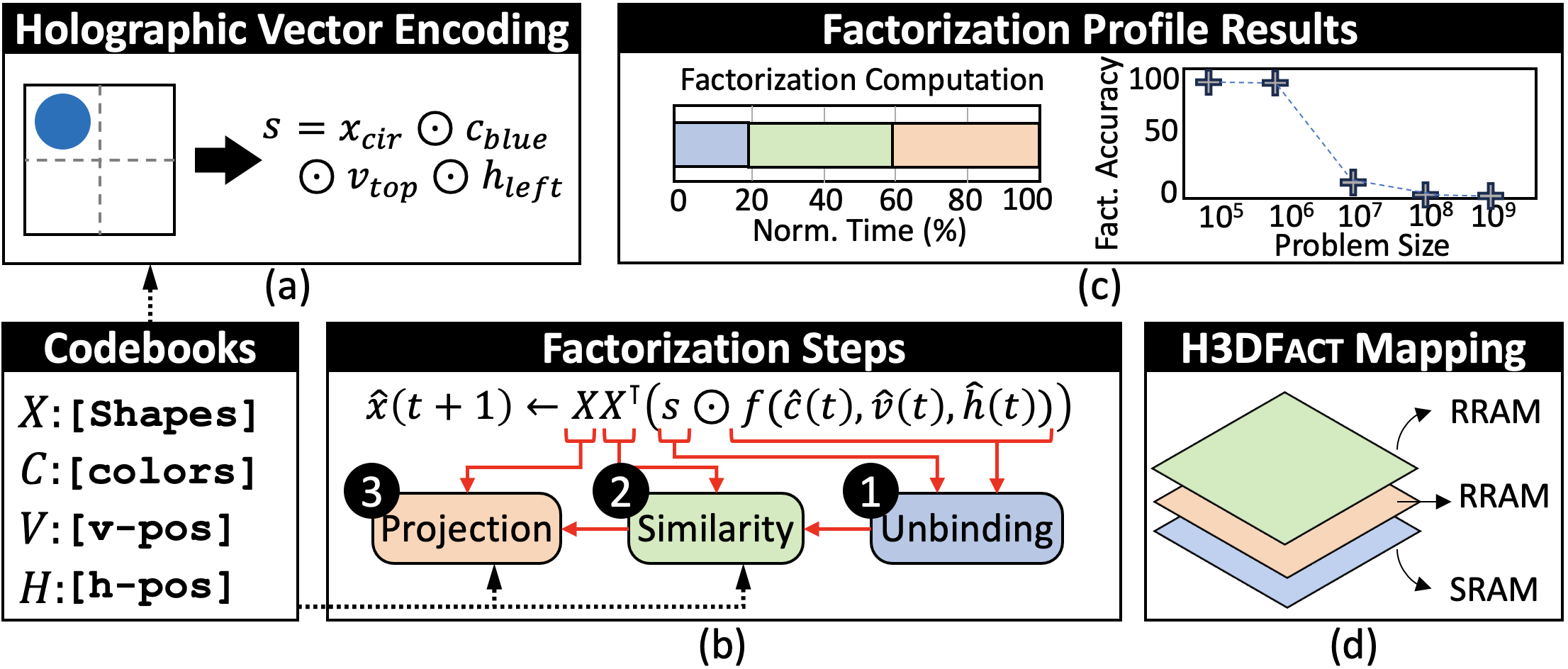}
    \caption{\textbf{Computational Primitives of the Holographic Vector Encoding and Factorization.} (a) Vector encoding of a visual object. (b) Algorithmic flow of the factorization problem. (c) Characterization results of the factorization operations. (d) An overview schematic of the proposed H3D integrated factorizer with hybrid-memory design.}
    \label{fig:operation}
\end{figure}

We illustrate here how holographic vectors are used to encode the compositional structure of objects and how the resonator network works to decode the contents of this structure via factorization. Consider an example of encoding visual objects, which are characterized by four attributes ($F=4$): shape, color, vertical position, and horizontal position. As demonstrated in Fig.~\ref{fig:operation}\textcolor{blue}{a}, each of these four attributes corresponds to a different $M$-sized codebook of randomly generated item vectors. This way, an object vector can be formed through the binding of vectors from these codebooks. 

The resonator network (factorization) works in the opposite direction. That is, it seeks to decompose an object vector into its constituent attribute vectors. The only inputs given to this algorithm are the composed object vector along with the individual codebooks of features. The algorithm compositionally searches through these codebooks to find the exact feature vectors. The following state-space equations describe this search (Fig.~\ref{fig:operation}\textcolor{blue}{b}): 

{\footnotesize
\[
\hat{x}(t+1) = g\big(XX^\top(s\odot \hat{c}(t)\odot\hat{v}(t)\odot\hat{h}(t)\big);\ \ X=[x_{cir}\ x_{tri}\ \ldots]
\]\[
\hat{c}(t+1) = g\big(CC^\top(s\odot \hat{x}(t)\odot\hat{v}(t)\odot\hat{h}(t)\big);\ \ C=[c_{blue}\ c_{red}\ \ldots]
\]\[
\hat{v}(t+1) = g\big(VV^\top(s\odot \hat{c}(t)\odot\hat{x}(t)\odot\hat{h}(t)\big);\ \ V=[v_{top}\ v_{bottom}]
\]\[
\hat{h}(t+1) = g\big(HH^\top(s\odot \hat{c}(t)\odot\hat{v}(t)\odot\hat{x}(t)\big);\ \ H=[h_{left}\ h_{right}]
\]
}

\noindent where $t$ is a time step; $s$ is the object vector; $\hat{x}$, $\hat{c}$, $\hat{v}$, and $\hat{h}$ hold the predicted values of the features $x$, $c$, $v$, and $h$, respectively. 

We observe that MVM operations dominate most of the computation time in the factorization algorithm. As shown in Fig.~\ref{fig:operation}\textcolor{blue}{c}, MVM operations within similarity and projection steps account for around 80\% of the total computation time. This result establishes a clear motivation for adopting a CIM design approach, which provides ways for MVM operations to always execute in a constant time irrespective of the problem size. 

Another motivation for using the CIM design approach is to address a major issue with the scaling of the factorization accuracy. Specifically, we observe a significant drop in the factorization accuracy with increasing the problem size (Fig.~\ref{fig:operation}\textcolor{blue}{c}). This accuracy drop is due to the limit cycle problem, which can be a limiting factor for large-scale factorization~\cite{langenegger2023memory}. One effective solution is to introduce stochasticity to break free of limit cycles and thus explore a substantially larger solution space. CIM devices are inherently stochastic; therefore, they provide a natural way for implementing this solution.

\subsection{Heterogeneous 3D CIM Acceleration}
\label{subsec:h3d_motivation}
Prior 3D integrated hardware designs have mainly focused on accelerating CNNs~\cite{murali2023continuing}, Transformers~\cite{li2023h3datten}, or monolithic 3D integration~\cite{dutta2020monolithic}. In contrast, \textsc{H3DFact} tackles a different MVM workload that is heavily used in high-dimensional cognitive systems, and maps different components of this factorization workload to hybrid RRAM/SRAM memory tiers (Fig.~\ref{fig:operation}\textcolor{blue}{d}). Moreover, \textsc{H3DFact} provides flexibility in designing with hybrid technology nodes, thus leading to significant improvements in the compute density, energy efficiency, and silicon footprint compared to iso-capacity 2D designs.
\section{Compute-In-Memory Primitives}
\label{sec:cim}
This section first presents a detailed circuit-level view of \textsc{H3DFact} memory tiers, including RRAM (Sec.~\ref{subsec:rram_tier}) and digital-SRAM tiers (Sec.~\ref{subsec:sram_tier}), and then discusses the benefits of \textsc{H3DFact} inherent stochasticity to factorization convergence (Sec.~\ref{subsec:stochastic}).


\subsection{RRAM Tier}
\label{subsec:rram_tier}

\begin{figure*}
    \centering
    \includegraphics[width=\linewidth]{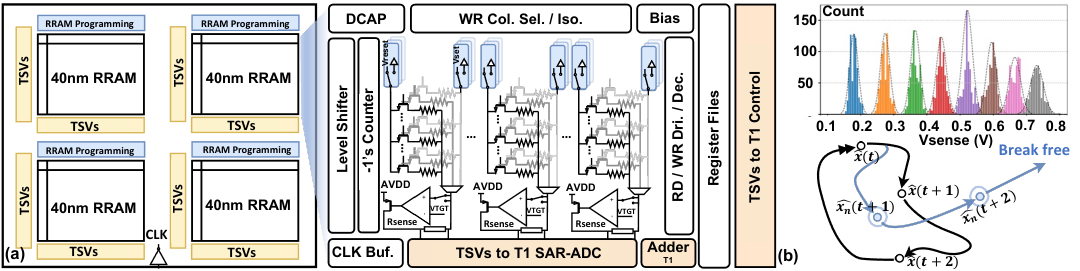}
    \caption{\textbf{\textsc{H3DFact} Array-Level Components.} (a) Legacy node RRAM tier-level view and building blocks for a single RRAM array. (b) The inherent stochasticity of \textsc{H3DFact} helps break limit cycles and benefit factorization convergence.}
    \label{fig:macro}
\end{figure*}

Fig.~\ref{fig:macro}\textcolor{blue}{a} provides a macro-level overview of the RRAM tier, depicting multiple arrays on a single tier. Each array is equipped with circuitry capable of executing MVM in the high-dimensional bipolar space ($\{-1,+1\}^D$). This circuitry includes a specialized -1's counter and an adder that processes bipolar quantities~\cite{spetalnick20232}. It is worth noting that the existing array designs for VSAs~\cite{karunaratne2021robust} often fall short of fully supporting the bipolar space, as they frequently map a bipolar element $\{-1,+1\}$ to a single-bit quantity, which is not suitable for the factorization algorithm that seeks to accumulate both positive and negative quantities within its computational flow.

The operation of the RRAM involves setting and resetting using high-voltage signals, necessitating the inclusion of isolation switches to protect peripherals against these high voltages~\cite{barkam2024memory}. Voltage regulation is achieved through a PMOS device connected to a power supply (AVDD) along with an operational amplifier (Fig.~\ref{fig:macro}\textcolor{blue}{b}). VTGT represents the target sensing voltage in the sensing path. Additionally, a current-sensing resistor (Rsense) is incorporated to enhance the process-voltage-temperature (PVT) immunity. Given that RRAM can be subject to frequent power-switching events, the design allows for different power-off modes (including a full shutdown) while enabling other tiers to remain active. These functions were experimentally validated using an RRAM chip fabricated with 40 nm technology~\cite{spetalnick202240nm}.

\subsection{Digital-SRAM Tier}
\label{subsec:sram_tier}
The interaction between the RRAM and the peripherals takes place through digital circuitry that includes an analog-to-digital (ADC) converter, an adder, and a controller (orange-colored blocks in Fig.~\ref{fig:macro}\textcolor{blue}{a}). One of the advantages of heterogeneous integration is the integration of systems with different technology nodes \cite{zhou2020near}. A potential area mismatch between RRAM and its peripherals results in MUX-sharing of the RRAM sensing \cite{spetalnick202240nm}. To fully unleash the system performance, digital components in \textsc{H3DFact} are designed in 16~nm advanced node, enabling a sensing path for each RRAM's output.

We adopt a hybrid-computing scheme for the frequently updated operation in the factorization as the write operation for RRAM is notorious for its humongous overhead \cite{yu2012switching}. The hybrid-computing scheme utilizes XNOR logic gates for bit-wise unbinding operation \cite{bankman2018always}. This is driven by constant memory write operation in unbinding updates for different time steps in factorization. 
In addition to the hybrid-computing scheme, a hybrid-memory (SRAM near-memory) scheme for buffering through-silicon vias (TSVs) data transfer in the H3D design, which will be further explained in Sec.~\ref{sec:3DArch}.
To reduce the TSV overheads, we only enable connections across different tiers in their input and output ports. For instance, connections only exist at the input row and output column for each RRAM array. This approach follows recent H3D design as excessive TSVs can severely damage not only the system-level PPA, but also RRAM ON/OFF ratio \cite{li2023h3datten}. However, this approach will require only one RRAM tier being activated. In Sec.~\ref{sec:3DArch}, we further discuss the architectural impact of RRAM activation. 
\subsection{Stochastic Factorizer}
\label{subsec:stochastic}
The unsupervised nature of the factorization's deterministic search could result in checking the same sequence of solutions multiple times across iterations, preventing convergence to the optimal solution in limited cycles. Inspired by \cite{langenegger2023memory}, one of the key insights is that the intrinsic stochasticity associated with memristive devices can substantially reduce the occurrence of such limit cycles. As shown in Fig.~\ref{fig:macro}\textcolor{blue}{b}, in-memory MVM readout results in a stochastic similarity vector with all the PVT variations aggregated altogether. The $\hat{x_n}(t+1)$ indicates noisy $\hat{x}$ at the $t+1$ time step. The hardware stochasticity enables the factorizer to break free of potentially being stuck at limited cycles and thus has the ability to explore a substantially larger space, demonstrating the potential to leverage device-level dynamics as a valuable source for application performance.
\section{\textsc{H3DFact} Architecture}
\label{sec:3DArch}
This section presents the \textsc{H3DFact} architecture, including the hardware design, workload mapping, and data flow (Sec.~\ref{subsec:3DArch_detail}), tier-to-tier interconnects (Sec.~\ref{subsec:tsv}), and floor plan scheme (Sec.~\ref{subsec:flooplann}).

\subsection{Proposed \textsc{H3DFact} Architecture}
\label{subsec:3DArch_detail}


\begin{figure}
    \centering
    \includegraphics[width=\linewidth]{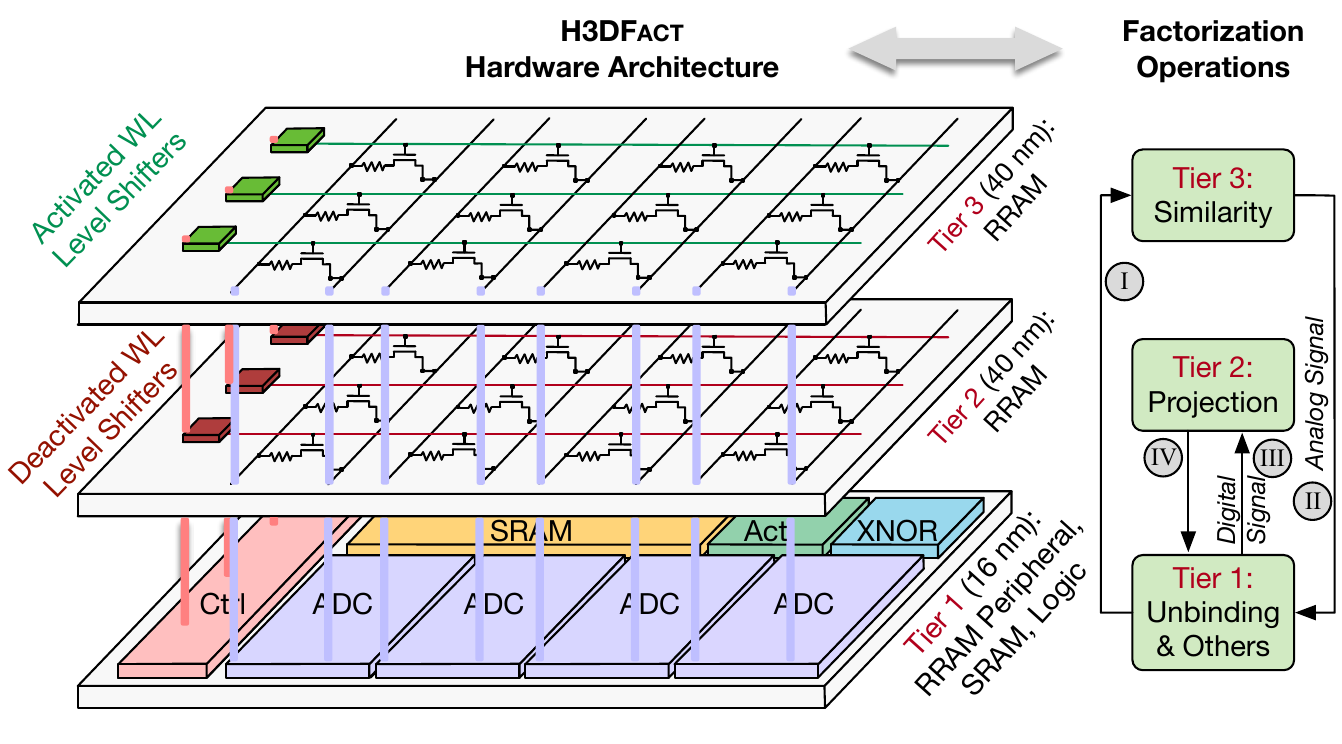}
    \caption{\textbf{\textsc{H3DFact} Architecture and Control Scheme.} The factorization computation kernels are partitioned among three vertical tiers. The control scheme for activating only one tier of RRAM CIM arrays when all RRAM tiers share the same vertical interconnects. Turning off the power to WL level shifters (red) will deactivate the current flow in the corresponding RRAM arrays.}
    \label{fig:h3dfact}
\end{figure}

\textbf{Factorization Workload Mapping.} \textsc{H3DFact} realizes factorization by partitioning its computational kernels into three tiers, in which similarity calculation, projection, digital operations, lie in tier-3, tier-2, and tier-1, respectively (Fig.~\ref{fig:h3dfact}). This design choice is related to the fact that the data is traversing in a digital or analog manner, where step I is the unbinding results for similarity calculation, step II is the similarity outputs that are represented with analog current, step III is the 4-bit digital result obtained from the similarity calculation, and step IV is the 1-bit digital data from projection. Analog data transfer from tier-3 to tier-1 and tier-2 to tier-1 is deemed to be negligible, as the analog current can flow through the TSV one-shot. On the other hand, the 4-bit digital similarity results obtained from tier-1 are passed to projection tier-2 to avoid multi-bit digital value transmission degrading system performance
Thus, \textsc{H3DFact} is designed with similarity at the top, projection at the middle, and digital circuits with advanced nodes at the bottom.

\textbf{Tier-2 \& Tier-3 RRAM CIM.} Inspired by~\cite{li2023h3datten}, a single set of RRAM peripherals is utilized collectively by both RRAM tiers (tier-2 and tier-3) in \textsc{H3DFact}, in a way that the interconnects from tier-1 link to every tier. Consequently, this architecture permits only one RRAM tier to be operational at any given time. This necessitates the inclusion of word line (WL) level shifters in each RRAM tier to manage their activation. Fig.~\ref{fig:h3dfact} depicts the control scheme of two tiers of RRAM. Due to the shared peripherals, the RRAM WLs, bit lines (BLs), and source lines (SLs) across the tiers are effectively connected vertically. To ensure that only a single RRAM tier is activated at a time, the design is equipped with a full shutdown capability where the non-active RRAM cells do not contribute to any column current.


\textbf{Tier-1 SRAM Digital Compute.}
We adopt SRAM in tier-1 to support greater-than-one factorization batch size. Considering a batch size of 100, after similarity calculation (tier-3), the similarity outputs propagate to tier-1 for analog-digital conversion. If without SRAM buffering, the tier-1 ADC-output signals are sent to tier-2 for projection calculation, which will violate single-RRAM tier activation because tier-3 is still computing similarity for data in the same batch. Therefore, we propose to adopt digital-SRAM in tier-1 to serve as buffers to support large batch factorization computation.

\textbf{Design Methodology Generalization.}
The \textsc{H3DFact} architecture is adept at handling the diverse parameters characteristic of resonator networks. Since resonator networks are parametrized with high dimensional vector dimensions $D$ and $F$ (Sec.~\ref{sec:background}), the \textsc{H3DFact} is configured with hardware dimensions to determine the number of rows of an RRAM array ($d$) and the number of RRAM subarrays of each tier ($f$). In this paper, we set $d=256$ and $f=4$ as an example of \textsc{H3DFact} design. 
This configuration not only accommodates the specified vector size, but also facilitates the parallel processing of multiple inputs by utilizing different subarrays.


\begin{table}[t!]
\renewcommand*{\arraystretch}{1.3}
\setlength\tabcolsep{3.5pt}
\centering
\caption{\textsc{H3DFact} Interconnect Specifications.}
\label{tab:tsv_spec}
\resizebox{\linewidth}{!}{%
\begin{tabular}{|c|c|c|c|c|c|}
\hline
\textbf{\begin{tabular}[c]{@{}c@{}}TSV\\ Diameter\end{tabular}} & \textbf{\begin{tabular}[c]{@{}c@{}}TSV\\ Pitch\end{tabular}} & \textbf{\begin{tabular}[c]{@{}c@{}}TSV Oxide\\Thickness\end{tabular}} & \textbf{\begin{tabular}[c]{@{}c@{}}TSV\\ Height\end{tabular}} & \textbf{\begin{tabular}[c]{@{}c@{}}Hybrid Bonding\\ Pitch\end{tabular}} & \textbf{\begin{tabular}[c]{@{}c@{}}Hybrid Bonding\\ Thickness\end{tabular}}\\ \hline
2~$\mu m$                                                      & 4~$\mu m$                                                      & 100~$nm$          & 10~$\mu m$          & 10~$\mu m$          & 3~$\mu m$          \\ \hline
\end{tabular}
}
\end{table}

\begin{figure}
    \centering
    \includegraphics[width=\linewidth]{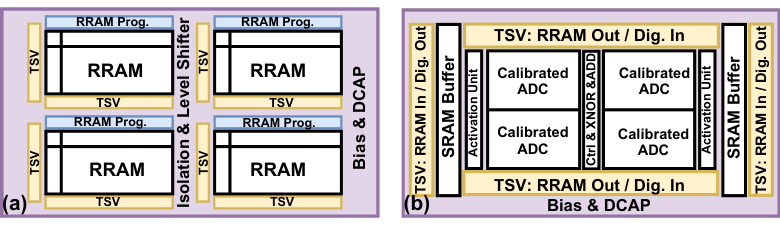}
    \caption{\textbf{\textsc{H3DFact} Floor Plan.} (a) RRAM tier-2/3. (b) Digital tier-1.}
    \label{fig:floorplan}
\end{figure}





\subsection{Tier-to-Tier Interconnects in \textsc{H3DFact}}
\label{subsec:tsv}

In Tab.~\ref{tab:tsv_spec}, we outline the parameters for the tier-to-tier interconnects in the \textsc{H3DFact} design. These assumptions are in line with recent H3D designs such as H3DAtten~\cite{li2023h3datten} and commercial designs such as AMD’s 3D V-Cache~\cite{swaminathan2021advanced}. For an RRAM array with $X$ rows and $Y$ columns, the total number of TSVs for connection to RRAM peripherals comprises $X$ for WLs, $Y$ for BLs, and $Y/2$ for SLs. Typically, larger arrays reduce TSV overhead but are less efficiently utilized than smaller ones. \textsc{H3DFact} opts to store the similarity matrix in the same array at each iteration to optimize TSV utilization.

Given the significant area costs associated with TSVs, our design strategy for \textsc{H3DFact} includes analog CIM and SAR-ADCs to minimize TSV area requirements. In analog CIM, the partial sums of MVM operations are conveyed as analog currents, requiring only a single set of interconnects to connect to an ADC in tier-1. \textsc{H3DFact} provides the flexibility to design RRAM peripheral circuitries in more advanced nodes~\cite{peng2020benchmarking,li2023h3datten}, thus we choose to assign each RRAM column with a 4-bit ADC. To validate, we quantize the similarity calculation to 4-bit, and observe no factorization accuracy drop while having even faster convergence than 8-bit ADC design (Sec.~\ref{subsec:eval_robust}).

\subsection{Floor Planning and Bonding of \textsc{H3DFact}}
\label{subsec:flooplann}
To verify that the tiers in the 3D stack of \textsc{H3DFact} are area-balanced and provide data for thermal analysis (Sec.~\ref{subsec:eval_thermal}), we conduct a floor plan approximation for each tier. The sizes of the CIM arrays and their associated peripherals are estimated using the calibrated NeuroSim framework~\cite{peng2020dnn+}, which has been cross-validated with actual RRAM-based CIM macros~\cite{spetalnick202240nm}.
Areas of other digital modules are extracted from the TSMC standard cell library.

Fig.~\ref{fig:floorplan}\textcolor{blue}{a} shows the floor plan of the \textsc{H3DFact} RRAM tier. Each RRAM subarray features a dimension of 256 $\times$ 256 size, with four subarrays designed for each tier. \textsc{H3DFact} can perform RRAM CIM operations in any particular subarray(s) by activating their corresponding WLs and BLs.

Fig.~\ref{fig:floorplan}\textcolor{blue}{b} shows the floor plan for the RRAM peripheral and SRAM digital compute tier. The memory controller and buffer are also placed at tier-1 to avoid many connections to other SoCs or packages, as the external pins and C4 bumps are on the bottom tier~\cite{murali2023continuing}. 
Regarding the bonding techniques between tier-to-tier TSVs, we consider a mix of face-to-face (F2F) and face-to-back (F2B). In F2B integration, TSVs bond multiple tiers. As TSVs penetrate through the silicon, the memory placement or the TSV usage gets restricted. While F2F does not pose any place and route restriction, it is impossible to integrate all three tiers using F2F integration \cite{murali2023continuing}, and a mix of F2F and F2B tier-to-tier connections are required for our three-tier \textsc{H3DFact} design. 

\section{\textsc{H3DFact} Evaluation}
\label{sec:eval}
This section evaluates \textsc{H3DFact} on factorization and holographic perceptual systems. We demonstrate that \textsc{H3DFact} achieves improved factorization accuracy, operational capacity (Sec.~\ref{subsec:eval_acc}), and hardware efficiency (Sec.~\ref{susbec:eval_hw}). We also analyze \textsc{H3DFact} thermal behavior (Sec.~\ref{subsec:eval_thermal}), illustrate the robustness of \textsc{H3DFact} with the RRAM silicon chip validation (Sec.~\ref{subsec:eval_robust}), demonstrate its role in visual perception task and discuss potential applications (Sec.~\ref{subsec:nvsa}).


\subsection{Accuracy and Operational Capacity}
\label{subsec:eval_acc}

\textbf{Accuracy Improvement.} Tab.~\ref{tab:acc} compares the factorization accuracy of \textsc{\textsc{H3DFact}} with baseline resonator network~\cite{frady2020resonator} under different number of attributes $F$ and code vectors $D$. While both baseline resonator network and \textsc{\textsc{H3DFact}} achieve 99\% accuracy under small $M^D$, it is well observed that \textsc{\textsc{H3DFact}} substantially enhances and maintains 99\% accuracy under high dimensionality, illustrating its improved scalability for larger factorization problem sizes. 

\textbf{Operational Capacity Improvement.} Tab.~\ref{tab:acc} also shows the number of iterations required to solve a given problem size with accuracy of at least 99\%. Compared with baseline resonator network~\cite{frady2020resonator}, \textsc{\textsc{H3DFact}} enables faster convergence and can solve problem sizes at five orders of magnitude larger at 99\% accuracy, illustrating \textsc{\textsc{H3DFact}} capable of lowering computational complexity with improved operational capacity. This observation is in line with CIM-based factorizer design~\cite{langenegger2023memory}.

\begin{table}[t!]
\centering
\caption{\textbf{Accuracy Evaluation.} Factorization accuracy and operational capacity comparison under different problem sizes.}
\renewcommand*{\arraystretch}{1.3}
\setlength\tabcolsep{3.5pt}
\resizebox{\linewidth}{!}{%
\begin{threeparttable}
\begin{tabular}{c|cccc|cccc}
\hline
\multirow{3}{*}{\begin{tabular}[c]{@{}c@{}}\\ \end{tabular}} & \multicolumn{4}{c|}{\textbf{Factorization Accuracy (\%)}}         & \multicolumn{4}{c}{\textbf{Number of Iterations\tnote{*}}}      \\ \cline{2-9} & \multicolumn{2}{c|}{$F$=3}               & \multicolumn{2}{c|}{$F$=4} & \multicolumn{2}{c|}{$F$=3}              & \multicolumn{2}{c}{$F$=4} \\ \cline{2-9}  & Baseline  & \multicolumn{1}{c|}{H3D}   & Baseline      & H3D      & Baseline  & \multicolumn{1}{c|}{H3D}  & Baseline    & H3D       \\ \hline
$D$=16                                                           & 99.4      & \multicolumn{1}{c|}{99.3}  & 99.2          & 99.2     & 4         & \multicolumn{1}{c|}{5}    & 31          & 33        \\
$D$=32                                                           & 99.3      & \multicolumn{1}{c|}{99.3}  & 99.1          & 99.2     & 13        & \multicolumn{1}{c|}{15}   & 234         & 140       \\
$D$=64                                                           & 99.1      & \multicolumn{1}{c|}{99.3}  & 89.9          & 99.2     & 43        & \multicolumn{1}{c|}{39}   & Fail         & 1347      \\
$D$=128                                                          & 96.9      & \multicolumn{1}{c|}{99.3}  & 0             & 99.2     & Fail      & \multicolumn{1}{c|}{108}  & Fail         & 17529     \\
$D$=256                                                          & 10.8      & \multicolumn{1}{c|}{99.2}  & 0             & 99.2     & Fail     & \multicolumn{1}{c|}{443}  & Fail         & 269931    \\
$D$=512                                                          & 0.2       & \multicolumn{1}{c|}{99.2}  & 0             & 99.2     & Fail     & \multicolumn{1}{c|}{1685} & Fail         & 2824079   \\ \hline
\end{tabular}
\begin{tablenotes}
        \item[*] \scriptsize Number of iterations required to reach at least 99\% accuracy under different problem sizes.
      \end{tablenotes}
      \end{threeparttable}
}
\label{tab:acc}
\end{table}


\subsection{Hardware Efficiency}
\label{susbec:eval_hw}

\begin{table*}[t!]
\huge
\centering
\caption{\textbf{Hardware Performance Evaluation.} Hardware resource and performance comparison between 2D and \textsc{H3DFact} Designs.}
\renewcommand*{\arraystretch}{1.4}
\setlength\tabcolsep{3pt}
\resizebox{\linewidth}{!}{%
\begin{tabular}{c|ccccccc|cccccc}
\hline
\multirow{3}{*}{\begin{tabular}[c]{@{}c@{}}\textbf{Design}\\ \textbf{Choice}\end{tabular}} & \multicolumn{7}{c|}{\textbf{Hardware Resource}}                                                                      & \multicolumn{6}{c}{\textbf{Performance}}                                                                                    \\ \cline{2-14} 
& \multicolumn{1}{c|}{\begin{tabular}[c]{@{}c@{}}Technology\\(RRAM)\end{tabular}} & \multicolumn{1}{c|}{\begin{tabular}[c]{@{}c@{}}Technology\\(RRAM Peripheral)\end{tabular}} & \multicolumn{1}{c|}{\begin{tabular}[c]{@{}c@{}}Technology\\(Digital)\end{tabular}} & \multicolumn{1}{c|}{\begin{tabular}[c]{@{}c@{}}Unbinding\\Operation\end{tabular}} & \multicolumn{1}{c|}{\begin{tabular}[c]{@{}c@{}}Similarity \& Projection\\Operation\end{tabular}} & \multicolumn{1}{c|}{\begin{tabular}[c]{@{}c@{}}ADC\\Count\end{tabular}} & \begin{tabular}[c]{@{}c@{}}TSV\\Count\end{tabular} & \multicolumn{1}{c|}{Area} & \multicolumn{1}{c|}{Frequency} & \multicolumn{1}{c|}{Throughput}  &\multicolumn{1}{c|}{\begin{tabular}[c]{@{}c@{}}Compute\\Density\end{tabular}}  & \multicolumn{1}{c|}{\begin{tabular}[c]{@{}c@{}}Energy\\Efficiency\end{tabular}} & \multicolumn{1}{c}{Accuracy} \\ \hline
\textbf{SRAM 2D} &   \multicolumn{1}{c|}{N/A}    &  \multicolumn{1}{c|}{N/A}   & \multicolumn{1}{c|}{16 nm}     &   \multicolumn{1}{c|}{SRAM Digital}      &   \multicolumn{1}{c|}{SRAM CIM}         &   \multicolumn{1}{c|}{0}                                                  & \multicolumn{1}{c|}{0 }    & \multicolumn{1}{c|}{0.114 mm$^2$}    &  \multicolumn{1}{c|}{200 MHz}    &  \multicolumn{1}{c|}{1.52 TOPS}         &    \multicolumn{1}{c|}{13.3 TOPS/mm$^2$}        &    \multicolumn{1}{c|}{50.1 TOPS/W}     &    95.8\%                                                                    \\
\textbf{Hybrid 2D} &  \multicolumn{1}{c|}{40 nm}     & \multicolumn{1}{c|}{40 nm}    &  \multicolumn{1}{c|}{40 nm}    &   \multicolumn{1}{c|}{SRAM Digital}      &   \multicolumn{1}{c|}{RRAM CIM}         &   \multicolumn{1}{c|}{1024}                                                  &  0   &  \multicolumn{1}{c|}{0.544 mm$^2$}   &   \multicolumn{1}{c|}{200 MHz}   &   \multicolumn{1}{c|}{1.52 TOPS}       &    \multicolumn{1}{c|}{2.8 TOPS/mm$^2$}        &  \multicolumn{1}{c|}{60.6 TOPS/W}       &    99.3\%                                                                     \\
\textbf{3-Tier H3D} &  \multicolumn{1}{c|}{40 nm}     &  \multicolumn{1}{c|}{16 nm}   &  \multicolumn{1}{c|}{16 nm}    &   \multicolumn{1}{c|}{SRAM Digital}      &  \multicolumn{1}{c|}{RRAM CIM}          &  \multicolumn{1}{c|}{1024}                                                   &  5120   &  \multicolumn{1}{c|}{0.091 mm$^2$}   &  \multicolumn{1}{c|}{185 MHz}    &    \multicolumn{1}{c|}{1.41 TOPS}       &   \multicolumn{1}{c|}{15.5 TOPS/mm$^2$}         &   \multicolumn{1}{c|}{60.6 TOPS/W}      &  99.3\%                                                                       \\ \hline
\end{tabular}
}
\label{tab:h3d_comparison}
\end{table*}

\textbf{Monolithic 2D Baseline Design Setup.} We evaluate the advantages of the proposed \textsc{H3DFact} by contrasting it with two distinct 2D architectures: a hybrid RRAM/SRAM design and an exclusively SRAM-based design (Tab.~\ref{tab:h3d_comparison}). For the hybrid 2D design, all modules are integrated using a 40~nm process node to accommodate the RRAM technology in a unified 2D structure. 
Conversely, the fully SRAM design scales all modules to the more advanced 16~nm nodes. We maintain identical computing resources and parameters across all these designs to ensure an equitable comparison. 

\textbf{Silicon Footprint Reduction.} Tab.~\ref{tab:h3d_comparison} shows that fully SRAM design in 2D requires an area of 0.114~mm$^2$ with all components in 16 nm. The 2D RRAM/SRAM hybrid design occupies up to 0.544~mm$^2$ despite involving no TSV overheads due to limitations in current RRAM fabrication technology. In contrast, the advanced node scaling and vertical integration in \textsc{H3DFact} allow a more compact footprint of 0.091~mm$^2$. Even accounting for all three tiers, \textsc{H3DFact} still provides appreciable reductions of 1.25$\times$ and 5.97$\times$ in total silicon cost compared to fully SRAM and hybrid 2D designs, respectively.

\textbf{Compute Density and Energy Efficiency Improvement.} Compared to 2D designs, \textsc{H3DFact} operates at a marginally lower frequency due to the parasitic capacitance introduced by TSVs and hybrid bindings thus resulting in a slight throughput penalty. Nevertheless, as in Tab.~\ref{tab:h3d_comparison}, \textsc{H3DFact} still demonstrates 1.2$\times$ higher compute density and 1.2$\times$ energy efficiency by scaling digital components and RRAM peripheral from 40 to 16 nm. Compared with the 2D fully deterministic digital SRAM baseline with all modules designed in 16~nm, \textsc{H3DFact} still achieves comparable energy efficiency with 5.5$\times$ higher compute density and 3.5\% higher factorization accuracy due to the associated intrinsic stochasticity (Fig.~\ref{fig:macro}\textcolor{blue}{c}).


\textbf{Compare with Other Factorization Accelerators.} Compared with recent PCM-based in-memory factorization~\cite{langenegger2023memory}, \textsc{H3DFact} achieves 1.78$\times$ throughput and 1.48$\times$ energy efficiency under the same silicon area by virtual of 3D stacking and improved compute density, with $>$99\% factorization accuracy.


\begin{figure}
    \centering
    \includegraphics[width=\linewidth]{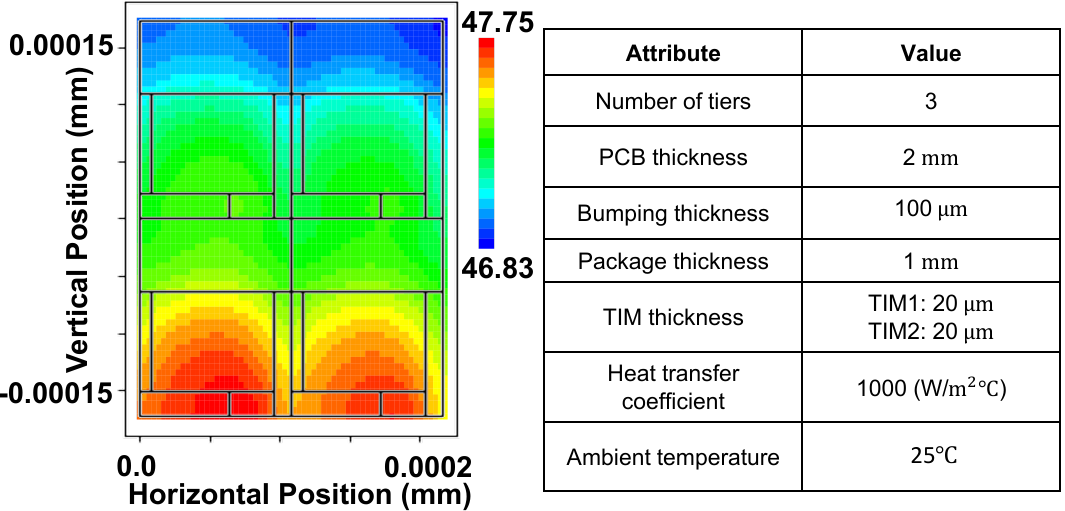}
    \caption{\textbf{Thermal Analysis.} Thermal map of \textsc{H3DFact} with its setup.}
    \label{fig:thermal}
\end{figure}


\subsection{Thermal Evaluation}
\label{subsec:eval_thermal}



\textbf{Thermal Analysis.} We utilize HotSpot~\cite{hotspot} tool to conduct thermal analysis of \textsc{H3DFact}, where we assign power densities to each component based on their respective floorplans (Fig.~\ref{fig:floorplan}). Our chip-level thermal setup includes hybrid bonds and TSVs to connect tiers 1-3, C4 bumps to connect tier 1 to the package, and thermal interface material (TIM) at the top for cooling. The parameters are summarized in Fig.~\ref{fig:thermal} and consistent with recent design~\cite{li2023h3datten}. As shown in Fig.~\ref{fig:thermal}, the tier temperatures for \textsc{H3DFact} range from 46.8~$^{\circ}$C to 47.8~$^{\circ}$C, where the 2D design is 44~$^{\circ}$C. With cooling being more effective at the center and high power density lying in the southern of each macro, as expected, there exist slight temperature increases toward the die southern region. Importantly, the 3D stacking approach used in \textsc{H3DFact} does not compromise the reliability of RRAM, as RRAM retention is adversely affected at temperatures exceeding 100~$^{\circ}$C~\cite{fang2010temperature}).


\subsection{Robustness Evaluation and Chip Validation}
\label{subsec:eval_robust}
\begin{figure}
    \centering
    \includegraphics[width=\linewidth]{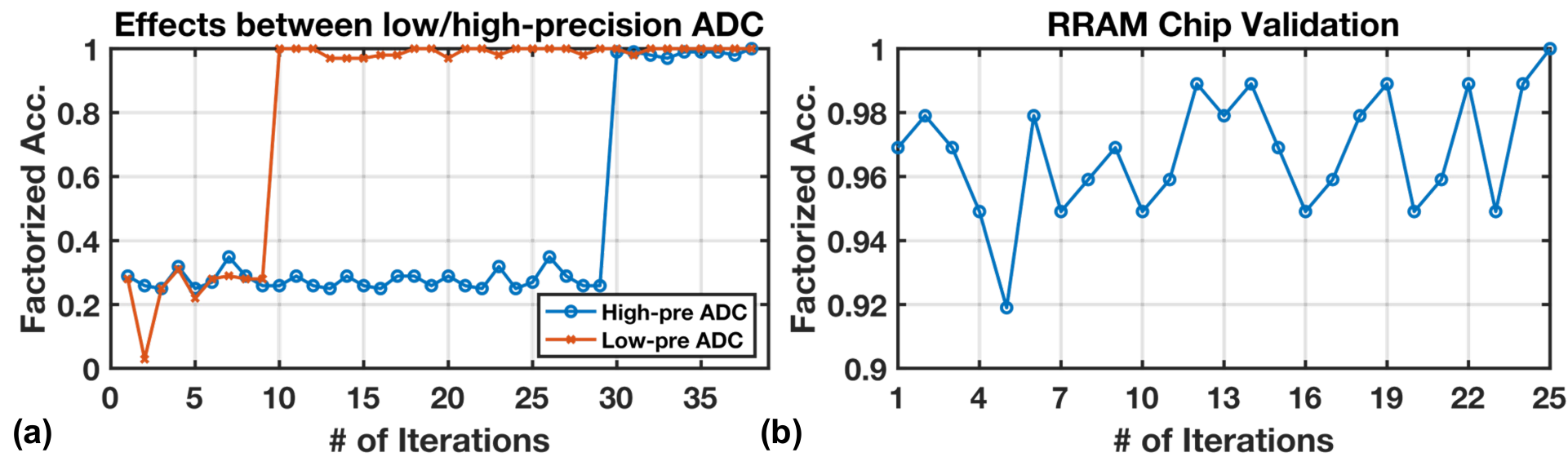}
    \caption{\textbf{Robustness Evaluation and Chip Validation.} (a) Factorization accuracy with low-precision (\textsc{H3DFact}) and high-precision ADC. (b) Factorization accuracy with 40~nm RRAM chip validation.}
    \label{fig:acc}
\end{figure}

\textbf{Convergence Speedup.} Lowering ADC precision can reduce hardware costs and enable faster convergence of holographic perceptual factorization while maintaining similar accuracy. As demonstrated in Fig.~\ref{fig:acc}\textcolor{blue}{a}, after applying low-precision 4-bit ADC to similarity calculation, the factorization converges to 99\% accuracy at 10th iteration, while it takes 30 iterations under 8-bit ADC. This is because lowering precision introduces quantization stochasticity, which prevents the factorizer stuck in a limit cycle and helps converge to the correct factorization in a shorter time (Fig.~\ref{fig:macro}\textcolor{blue}{c}).


\textbf{RRAM Testchip Validation.}
We validate the effectiveness of our proposed \textsc{H3DFact} design on the fabricated 40~nm RRAM testchips \cite{spetalnick202240nm,spetalnick20232}. We extract inherent noise parameters from RRAM testchips by measuring the readout signal and incorporate their statistics into the developed holographic perceptual factorization framework. We also adjust the threshold value accordingly as the designed readout peripheral is able to change the readout voltage ($V_{TGT}$ in Fig.~\ref{fig:macro}). As shown in Fig.~\ref{fig:acc}\textcolor{blue}{b}, the RRAM testchip validated \textsc{H3DFact} design achieves $>96\%$ factorization accuracy at one-shot and reaches 99\% accuracy after 25 iterations.

\subsection{Holographic Perception Task Evaluation}
\label{subsec:nvsa}

\begin{figure}
    \centering
    \includegraphics[width=\linewidth]{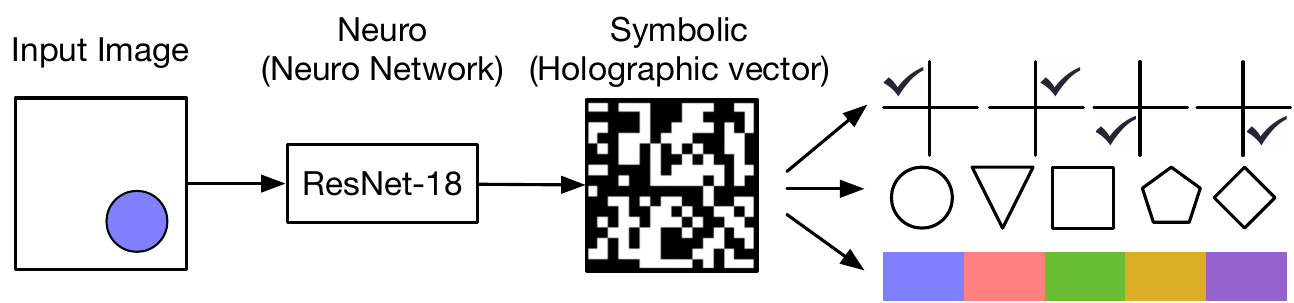}
    \caption{\textbf{Holographic Neuro-Symbolic Evaluation.} The visual perception task involves neural networks for feature mapping and holographic vectors for attribute reasoning.}
    \label{fig:raven}
\end{figure}

\textbf{Holographic Perception Accuracy.} Fig.~\ref{fig:raven} demonstrates the role of \textsc{H3DFact} in visual perception task to disentangle the attributes of raw images. The system consists of two components: a neural network to map input images to holographic perceptual vectors, and \textsc{H3DFact} to disentangle the approximate product vector using a known set of image attributes (e.g., type, size, color, and position). Evaluated on the relational and analogical visual reasoning (RAVEN) dataset~\cite{zhang2019raven}, \textsc{H3DFact} achieves 99.4\% accuracy of attributes estimation.



\textbf{Extensible to Other Applications.} \textsc{H3DFact} is effective beyond visual perception, as factorization plays a fundamental role in perception and cognition (e.g., analogical reasoning, tree search, and integer factorization). This hierarchical cognition capability can be potentially applied in autonomous systems where robustness is critical~\cite{wan2022analyzing,hsiao2023silent,wan2023berry}.
We envision \textsc{H3DFact} paves the way for solving complex combinatorial search and hierarchical cognition problems in next-generation neuro-symbolic AI systems.


\section{Conclusion}
\label{sec:conclusion}
\textsc{H3DFact} is the first H3D integrated CIM design unlocking efficient and scalable high-dimensional holographic vector factorization. \textsc{H3DFact} exploits the computation-in-superposition capability, emerging memory technologies, and intrinsic hardware stochasticity, and consistently improves factorization accuracy and operational capacity, with 5.5$\times$ compute density, 1.2$\times$ energy efficiency, and 5.9$\times$ less silicon footprint compared to iso-capacity 2D designs. We envision \textsc{H3DFact} being useful in exploring other robust and efficient cognitive and neuro-symbolic AI systems.

\section*{Acknowledgements}
The authors thank Wantong Li, Gauthaman Murali, and Lingjun Zhu (Georgia Tech) for helpful discussions. This work was supported in part by CoCoSys, one of seven centers in JUMP2.0, a Semiconductor Research Corporation (SRC) sponsored by DARPA.

\bibliographystyle{IEEEtran}
\bibliography{refs}
\end{document}